\documentclass[a4paper,fleqn,usenatbib]{mnras}

\usepackage[T1]{fontenc}
\usepackage{ae,aecompl}

\usepackage{graphicx}	
\usepackage{amsmath}	
\usepackage{amssymb}	

\def\msun{{\rm\,M_\odot}}

\def\gtsima{$\; \buildrel > \over \sim \;$}
\def\simgt{\lower.5ex\hbox{\gtsima}}

\def\msun{\hbox{M$_\odot$}}
\title[Milky Way open cluster candidates]{Assessing the physical reality of Milky Way open cluster candidates}

\author[A.E. Piatti et al.]{Andr\'es E. Piatti$^{1,2,3}$\thanks{E-mail: andres.piatti@unc.edu.ar},
Denis M.F. Illesca$^{1,2,3}$, 
Agustina A. Massara$^{3}$, 
Mat\'{\i}as Chiarpotti$^{3}$,
\newauthor Daiana Rold\'an$^{3}$, 
Micaela Mor\'on$^{3}$, 
and Fabrizio Bazzoni$^{3}$\\
$^{1}$Instituto Interdisciplinario de Ciencias B\'asicas (ICB), CONICET-UNCUYO, 
Padre J. Contreras 1300, M5502JMA, Mendoza, Argentina\\
$^{2}$Consejo Nacional de Investigaciones Cient\'{\i}ficas y T\'ecnicas, Godoy Cruz 
2290, C1425FQB,  Buenos Aires, Argentina\\
$^{3}$Facultad de Ciencias Exactas y Naturales, Universidad Nacional de Cuyo,
Padre J. Contreras 1300, M5502JMA, Mendoza, Argentina\\
}

\date{Accepted XXX. Received YYY; in original form ZZZ}

\pubyear{2022}

\begin{document}
\label{firstpage}
\pagerange{\pageref{firstpage}--\pageref{lastpage}}
\maketitle

\begin{abstract}
We report results on the analysis of  eleven new Milky Way open cluster candidates, 
recently discovered from the detection of stellar overdensities in the
Vector Point diagram, by employing extreme deconvolution Gaussian mixture models.
We treated these objects as real open clusters and derived their
fundamental properties with their associated intrinsic dispersions by exploring the parameter
space through the minimization of likelihood functions on
generated synthetic colour-magnitude diagrams (CMDs). The intrinsic dispersions of
the resulting ages turned out to be much larger than 
those usually obtained for open clusters. Indeed, they resemble those of ages and
metallicities of composite star field populations.  We also traced their stellar number density
profiles and mass functions, derived their total masses, Jacobi and tidal radii, 
which helped us as criteria while assessing their physical nature as real open clusters.
Because the eleven candidates show a clear gathering of stars in the proper motion plane
and some hint for similar distances, we concluded that they are  possibly
sparse groups of stars.
\end{abstract} 




\section{Introduction}
Tons of data have been made available from sky surveys
that have allowed us to embark in searches for new open
cluster candidates \citep{ivanovetal2017,torrealbaetal2019},
among them, {\it Gaia} \citep{gaiaetal2022b}; 
SMASH \citep{nideveretal2021}; VVV \citep{minnitietal2010}, etc.
The background motivation for such an endeavour comprises,
among others, the knowledge of the Milky Way open cluster system
\citep{diasetal2021w}; the recovery of the Milky Way
disc star formation history \citep{andersetal2021}; 
the trace of Galactic spiral arm sub-structures
\citep{cantatgaudinetal2020}, etc.
To explore and exploit such a giant data volume,
automatic computer-based engines have been developed
\citep{hr2021,castroginardetal2021}. These search engines
deal with the recognition of overdensities of stars in an
$N$-dimensional space, including proper motions,
parallaxes, photometric properties, sky positions,
etc, as independent dimensions (variables). Their
success in identifying new open cluster candidates
has varied depending on the engines' complexity.

Recently, \citet{jaehnigetal2021} applied extreme deconvolution
gaussian mixture models on {\it Gaia} Data Release 2
proper motions and parallaxes \citep{gaiaetal2016,gaiaetal2018b} and identified 
11 previously uncovered Vector Point diagrams' overdensities,
which occupy a compact volume in the proper motion space.
The stars populating these overdensities are distributed
in the colour-magnitude diagram (CMD) resembling those
of open clusters. Based on this similarity, they 
performed theoretical isochrone fits and derived representative
reddening, ages, and distances. Nevertheless, they stressed
the status of these 11 objects as open cluster candidates and
made clear that detailed analyses are needed in order to confirm them as bonafide
open clusters. They called them XDOCC
(eXtreme Deconvolution Open Cluster Candidates) and
numbered them from $\#$1 to $\#$11.

\begin{figure*}
\includegraphics[width=\textwidth]{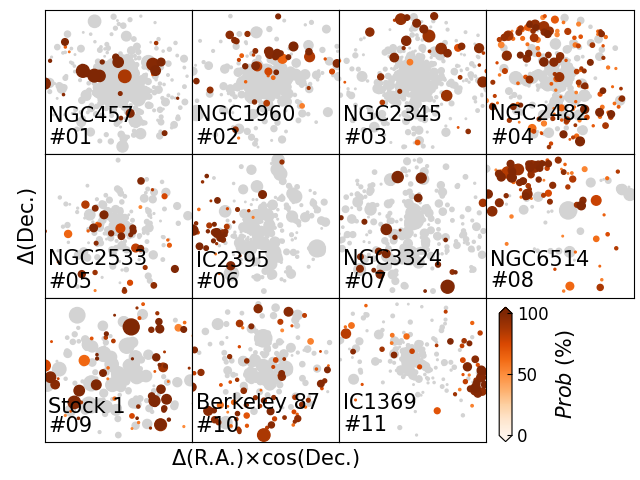}
\caption{Sky charts of stars with assigned membership probability (Prob) higher than
50 per cent projected on the field of known open clusters (gray symbols) and XDOCCs
(coloured symbols), respectively. The size of the symbols is proportional to the star brightness.
Each panel, centred on the known open cluster, indicates its name 
and the number of the XDOCC.}
\label{fig1}
\end{figure*}

Looking at their stellar distributions in the Galactic coordinate system 
\citep[see Figures 9-11 in][]{jaehnigetal2021}, most of the XDOCCs do not show the 
expected \citet{king62}  profile \citep{piskunovetal2007,kharchenkoetal2013}, but stars scattered across the analyzed field.
This appearance led us to remind that field stars may also distribute in the CMD
giving the appearance of the star sequence of an open cluster, as \citet{bm1973} shown. Therefore,
the presence of star sequences in CMDs should not be taken as a direct proof 
of the existence of a real open cluster. Relatively faint field star 
Main Sequences, for example, can mimic the lower part of open cluster Main Sequences.
Precisely, the aim of this work consists in revisiting the 11 XDOCCs and 
providing with an assessment on their nature as genuine open clusters. 
The present results highlight the importance of considering not only the proper motions
distribution to conclude on the existence of a physical system,
but also, its spatial distribution, size, mass, etc. 

In Section 2 we estimate XDOCC fundamental
parameters disentangling observational errors from the intrinsic dispersion and
compare the resulting values with those derived by \citet{jaehnigetal2021}.  We
also discuss different astrophysical aspects that arise from
considering the available data to conclude on the unsupported existence
of new open clusters. In Section 3 we summarize the main conclusions of this
work and suggest some sanity check analysis for future searches of open cluster candidates.

\section{Data analysis and discussion}

For comparison purposes with the work by  \citet{jaehnigetal2021},
we employed their same data sets, which include stars with assigned membership probabilities 
higher than 50 per cent. Since the 11 XDOCCs were identified projected on the field
of 11 known open clusters, we also retrieved the same information for them.
We started by constructing schematic sky charts for the 11 XDOCCs, which we depict
in Fig.~\ref{fig1}. Stars of the known open clusters were drawn with light gray filled circles 
and those of the  11 XDOCCs with coloured ones, according to their membership
probability. We represented the stars with circles of different size, which are proportional 
to the star brightness. Each panel is centred on the known open cluster and its size is such
that it includes all the retrieved stars in the field. At first glance, the known open clusters are 
clearly visible, with the sole
exception of NGC~6514 (middle-right panel), which is a diffuse nebula known as the "Trifid" Nebula
\citep{gk1984}. As for the 11 XDOCCs, their appearance do not seem to resemble
those of real open clusters, with some exceptions. Indeed, although open clusters 
can have stars spatially sparsely distributed in comparison with those in
globular clusters, they all have a core (central) region of higher stellar density. 
Open clusters also usually show brighter stars more centrally concentrated than
fainter stars. Based on these qualitative descriptions of an open cluster, we distinguished 
those XDOCCs with a chance of being real stellar aggregates from those that seem
more probably to be the result of a superposition of stars aligned along the line-of.sight.
We included such a classification in Table~\ref{tab1} with an Y or N, respectively.

\subsection{Stellar mass functions}

\begin{figure*}
\includegraphics[width=\textwidth]{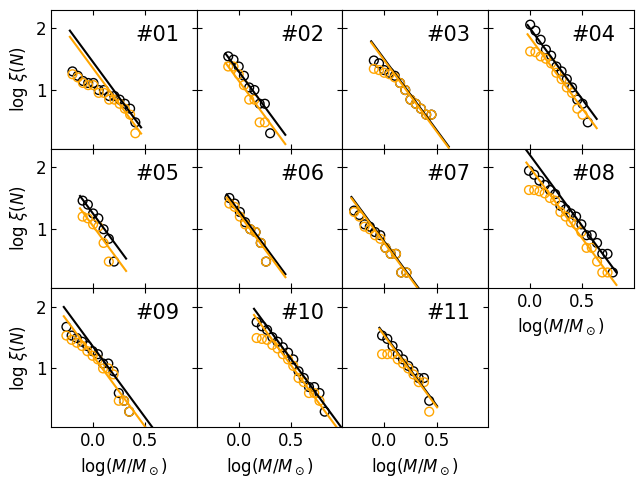}
\caption{Mass function of the XDOCCs, with the corresponding 
numbers labeled in the respective panels. Black and orange circles correspond
to mass functions built from stars with membership probabilities higher than
50 and 90 per cent, respectively. The coloured straight lines match the 
respective mass distributions.}
\label{fig2}
\end{figure*}

\begin{figure*}
\includegraphics[width=\textwidth]{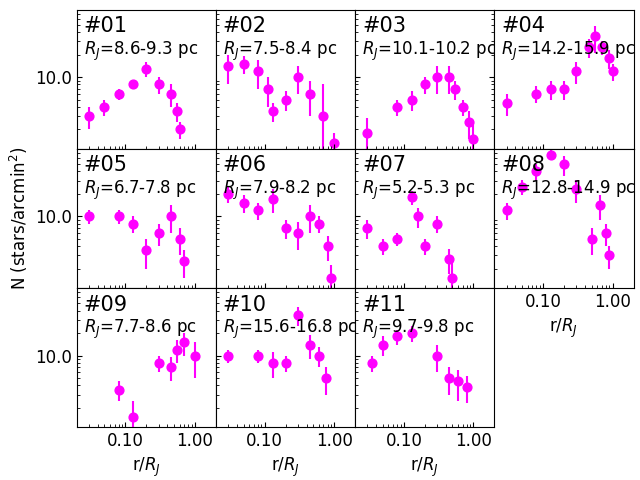}
\caption{Stellar number density profiles of the XDOCCs, with the corresponding 
numbers labeled in the respective panels. The derived Jacobi radius ranges are also
indicated (see text for details).}
\label{fig3}
\end{figure*}

\citet{joshietal2016} derived a relationship between open cluster mass and size that
we used to probe the reality of the 11 XDOCCs as physical systems. By assuming
that they are open clusters, we first constructed
their mass functions from which we estimated an upper limit for their initial total masses.
In order to do that, we used stars with membership probabilities higher than 50 and 90
per cent, respectively, with the aim of evaluating mass function uncertainties. While
counting the number of stars per mass interval, we adopted a mass bin of 
log($M$ /$\msun$) = 0.05. The individual stellar masses were interpolated using the
theoretical isochrones computed by  
\citet[][PARSEC v1.2S\footnote{http://stev.oapd.inaf.it/cgi-bin/cmd}]{betal12}, the
{\it Gaia} $G$ magnitudes and the ages, distances and interstellar absorptions 
$A_V$ derived for the 11  XDOCCs by \citet{jaehnigetal2021}.
The resulting mass functions are shown in  Fig.~\ref{fig2},
where we distinguished with black and orange open clusters those built for stars with
membership probabilities higher than 50 and 90 per cent, respectively. As can be seen,
there is only small differences between them. We then matched on them a \citet{kroupa02}'s 
mass function profile, which in turn  we used to compute the total mass for stars more massive 
 than  0.5$\msun$. We then computed
the XDOCCs' Jacobi radii using the expression \citep{spitzer1987}:

\begin{equation}
R_J = \left(\frac{M_{XDOCC}}{3 M_{MW}}\right)^{1/3}\times R_{GC}
\end{equation}

\noindent where $M_{XDOCC}$ is the XDOCC's mass derived above and $M_{MW}$ is the Milky Way 
mass comprised within a radius equal to the XDOCC's Galactocentric distance ($R_{GC}$). 
The $R_{GC}$ values were taken from \citet{jaehnigetal2021}, while $M_{MW}$ values
were interpolated in the Milky Way mass versus Galactocentric relationship obtained by 
\citep{birdetal2022}. We obtained two different $R_J$ values using the  derived 
XDOCC's mass estimates  for membership probabilities higher than 50 and 90
per cent, respectively. We included these values as labels in the different panels
of Fig.~\ref{fig3}.

\begin{figure}
\includegraphics[width=\columnwidth]{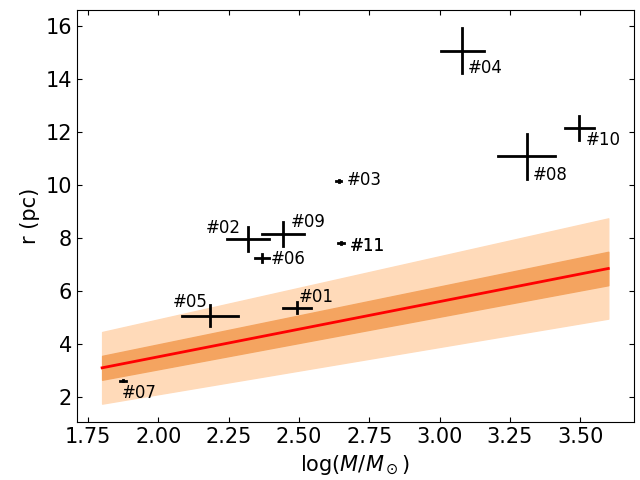}
\caption{Relationship between radii and masses of the 11 XDOCCs. The red line,
and orange and light-red shaded regions around it represent the relationship
derived by \citet{joshietal2016} and the 1$\sigma$ and 3$\sigma$ confidence
intervals, respectively. The numbers of the respective XDOCCs are also
indicated.}
\label{fig4}
\end{figure}

The resulting  $M_{XDOCC}$ and $R_J$  ranges were plotted 
in Fig.~\ref{fig4} as horizontal and vertical segments, respectively, where we also
included the radius versus mass relationship obtained by \citet{joshietal2016}. 
We used the locci of the XDOCCs in Fig.~\ref{fig4}  as an additional
criterion for assessing on their physical reality as open clusters. Particularly, objects that fall
within a 3$\sigma$ confidence interval (XDOCC~01, 05, and 07) were considered as possible
real systems, and for them, we included an Y in Table~\ref{tab1}.

\subsection{Stellar number density profiles}

In order to quantify the visual appearance of the 11 XDOCCs (see Fig.~\ref{fig1}), we 
built their stellar number density profiles. To to this, we counted the number of stars in annuli
centred on the XDOCCs of width 0.050, 0.067, 0.100, and 0.200 times their Jacobi radii 
 (for membership probabiilties higher than 90 per cent), respectively,
and then averaged all the obtained density values,  previously rebinned from 
interpolation, and computed their standard errors. The
resulting binned number density profiles are shown in Fig.~\ref{fig3},   where we
used  the ratio between the distance to the XDOCC's centre and the Jacobi radius
for comparison purposes.
In addition, dealing with $R_J$   values allow us to evaluate the
level of star mass segregation and to probe whether there are unbound stars among those 
identified as  XDOCCs'  members \citep{kupperetal2010b}. The latter is an interesting
aspect to analyze given the sparse appearance of the 11  XDOCCs.

The stellar radial profile of an open cluster
is expected to follow a King's \citep{king62} model
\citep[][and references therein]{piskunovetal2007,kharchenkoetal2013}, as follows:

\begin{equation}
  f(r)  = k\times\left( \frac{1}{\sqrt{1+(r/r_c)^2}} - \frac{1}{\sqrt{1+(r_t/r_c)^2}} \right)^2,
\end{equation}  

\noindent where $k$ is a constant, $r_c$ and $r_t$ are the core and tidal radius, respectively.
 Eq. (2) implies that there is  not any clusters' members beyond $r_t$. Likewise,  $r_t$ 
cannot be larger than the derived  $R_J$. Recently, \citet{zhongetal2022} showed that a two-components model with a 
King core distribution and a logarithmic Gaussian outer halo
distribution describe better the  internal and external structural features of open clusters.
They found that core, half-mass, tidal and Jacobi radii are statistically linearly related, 
which suggests  that the inner and outer regions of the clusters are interrelated and 
follow similar evolutionary processes. Because none of the XDOCCs' number density
profiles of Fig.~\ref{fig3} resembles a King's \citep{king62} profile, we concluded 
that the XDOCCs are not real open clusters, and included
an N in the fourth column of Table~\ref{tab1}. Note that
the remarkable drop of the stellar  number  density  profiles towards the inner regions 
is not caused by crowding effects, because XDOCCs are composed by a relatively sparse 
group of stars. 

\begin{table*}
\caption{Assessments on the reality of XDOCCs as open clusters. }
\label{tab1}
\begin{tabular}{lccccccc}\hline\hline
Name 	     & sky chart & size and mass  & radial profile  & CMD & adopted \\\hline
XDOCC-01    &     N       &		Y         &		N		 &   	N	& N	\\
XDOCC-02    &      N        &	N	      &		N		 &   	N	& N\\
XDOCC-03    &       N       &	N	      &		N		 &   	N	& N	\\
XDOCC-04    &       N       &	N	      &		N		 &   	N	& N	\\
XDOCC-05    &        N      &	Y	      &		N		 &   	N	& N	\\
XDOCC-06    &        Y      &	N	      &		N		 &   	N	& N	\\
XDOCC-07    &        N      &	Y	      &		N		 &   	N	& N	\\
XDOCC-08    &         Y      &	N	      &		N		 &   	N	& N	\\
XDOCC-09    &         N      &	N	      &		N		 &   	N	& N	\\
XDOCC-10    &          N     &	N	      &		N		 &   	N	& N	\\
XDOCC-11    &          Y     &	N	      &		N		 &   	N	& N	\\\hline
\end{tabular}
\end{table*}




\subsection{Colour-magnitude diagrams}

\begin{figure*}
\includegraphics[width=\textwidth]{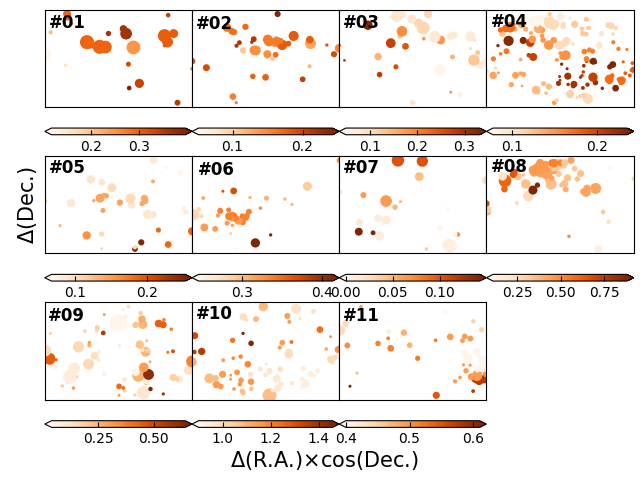}
\caption{Sky charts of XDOCCs' stars (see Fig.~\ref{fig1}); the size
of the symbols is proportional to the star brightness. Stars are
coloured according the its $E(B-V)$ value (mag) as shown in the 
respective colour bar.}
\label{fig5}
\end{figure*}

\begin{figure*}
\includegraphics[width=\textwidth]{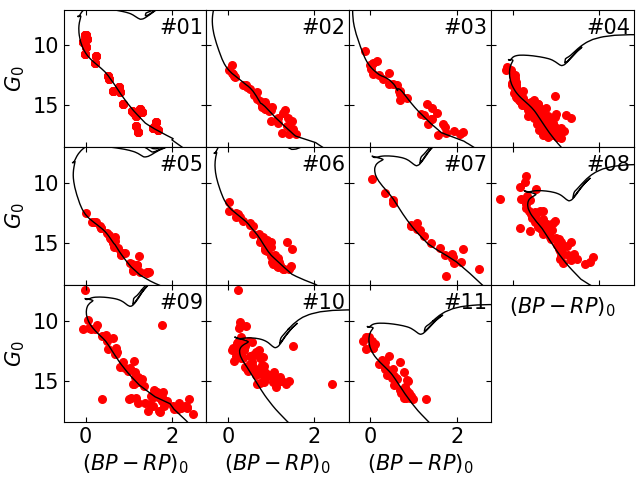}
\caption{Reddening corrected {\it Gaia} CMDs of the 11 XDOCCs as labelled in 
the upper-right corner of the panels. The theoretical isochrone of \citet[][PARSEC v1.2S]{betal12} 
corresponding to the parameters given in Table~\ref{tab2} and for the mean observed 
parallaxes derived by \citet{jaehnigetal2021} are superimposed.}
\label{fig6}
\end{figure*}

We first obtained individual stellar reddenings through the 
GALExtin\footnote{http://www.galextin.org/} interface \citep{amoresetal2021}, by using the 
Milky Way reddening map of  \citet{chenetal2019}, which was built specifically for 
{\it Gaia} bandpasses. We then corrected the $G$ magnitudes and $BP-RP$  colours using the 
retrieved total absorptions ($A_G$) and the total to selective absorption ratios given by 
\citet{chenetal2019}. The absorption uncertainties $\sigma$($A_G$) span
from 0.003 up to 0.020 mag, with an average of 0.010 mag at any $A_G$ interval.
Fig.~\ref{fig5} illustrates the spatial reddening variations for the XDOCCs' stars,
with stars coloured according to their $E(B-V)$ values, while
Fig.~\ref{fig6} depicts the reddening corrected CMDs.

We  performed isochrone fits to the reddening corrected XDOCCs' CMDs, 
not assuming a solar metallicty \citep{jaehnigetal2021}, but including the metallicity ([M/H]),
the age, the total mass, and the binary fraction as free  parameters. 
The fundamental parameters were derived by employing specific routines of the 
Automated Stellar Cluster Analysis code \citep[\texttt{ASteCA,}][]{pvp15},
which is able to derive them simultaneously. \texttt{ASteCA} relies on the construction
of a large number of synthetic CMDs from which it finds the one which best resembles
the observed CMD. Thus, the metallicity, the age, the distance, the reddening, the star 
cluster present mass and the binary fraction associated to that best representative
generated synthetic CMD are  adopted as the best-fitted star cluster properties. 

\texttt{ASteCA} is able to handle with a wide range of values of the aforementioned 
parameters. However, since the stars used share similar parallaxes 
\citep[see Table~1 and Figures 9-11 in][]{jaehnigetal2021}, we constrained the generation of 
synthetic CMDs to those with distance modulus around the mean observed parallaxes.
We fitted theoretical isochrones  computed by 
\citet[][PARSEC\footnote{http://stev.oapd.inaf.it/cgi-bin/cmd}]{betal12} for the
{\it Gaia} DR2 photometric system. Particularly, we chose PARSEC v1.2S isochrones spanning 
metallicities ($Z$ = 0.0152$\times$10$^{\rm [Fe/H]}$) from 0.003 dex up to 0.038 dex, in steps of 0.001 dex 
and log(age /yr) from 7.0 up to 9.0 in steps of 0.025. Because photometric errors are
not include in Table~2 of \citet{jaehnigetal2021}, we interpolated them from Figures~9-11
in \citet{evansetal2018}. To derive the errors in the {\it Gaia} colour, we added in quadrature
those of the individual involved magnitudes. We note that reliable photometric uncertainties are needed
to uncover the intrinsic dispersion in the CMD. If points in the CMD are used without errors 
then, the observed scatter is considered as the result of the combined
intrinsic dispersions of the fundamental parameters (reddening, distance, age, etc). 
When photometry errors are taken into account, then the resulting astrophysical
properties uncertainties are better constrained. In practice, the intrinsic dispersion
should be smaller than observed photometric ones.

\texttt{ASteCA} generates synthetic CMDs by adopting the initial mass function given by 
\citet{kroupa02} and a minimum mass ratio for the generation of binaries of 0.5. The total
observed star cluster mass and its binary fraction were set in the ranges 100-5000 $\msun$ 
and 0.0-0.5, respectively. In brief, \texttt{ASteCA} explores the parameter space of the 
synthetic CMDs through the minimization of the likelihood function defined by 
\citet[][the Poisson likelihood ratio (eq. 10)]{tremmeletal2013} using a parallel tempering 
Bayesian MCMC algorithm, and the optimal binning \citet{knuth2018}'s method.
The uncertainties associated to the derived parameters are estimated from the standard bootstrap 
method described in \citet{efron1982}. We refer the reader to \citet{pvp15} where details 
related to the implementation of these algorithms are provided. The resulting fundamental parameters 
are listed in Table~\ref{tab2} and the object CMDs with the isochrone corresponding to those
parameter values are illustrated in Fig.~\ref{fig6}.

\begin{table*}
\caption{\texttt{ASteCA} results for XDOCC objects}
\label{tab2}
\begin{large}
\begin{tabular}{lcccc}\hline\hline
Name 	  & log($t$ /yr)       & [Fe/H]         & Mass         & binary        \\
           &                    & (dex)          & ($\msun$)    &fraction      \\\hline
XDOCC-01   &   8.13$\pm$0.28   &  0.40$\pm$0.20 & 213$\pm$89   & 0.38$\pm$0.14 \\
XDOCC-02   &   7.47$\pm$0.43   &  0.24$\pm$0.18 & 343$\pm$122  & 0.45$\pm$0.12 \\
XDOCC-03   &   7.23$\pm$0.48   &  0.25$\pm$0.24 & 189$\pm$93   & 0.17$\pm$0.13 \\
XDOCC-04   &   8.47$\pm$0.06   &  0.23$\pm$0.12 & 1543$\pm$328 & 0.32$\pm$0.10 \\
XDOCC-05   &   7.73$\pm$0.45   &  0.17$\pm$0.19 & 188$\pm$126  & 0.22$\pm$0.15 \\
XDOCC-06   &   7.67$\pm$0.40   &  0.39$\pm$0.15 & 229$\pm$123  & 0.41$\pm$0.12 \\
XDOCC-07   &   8.76$\pm$0.31   &  0.24$\pm$0.24 & 125$\pm$38   & 0.43$\pm$0.13 \\
XDOCC-08   &   8.75$\pm$0.12   &  0.34$\pm$0.12 & 661$\pm$304  & 0.40$\pm$0.12 \\
XDOCC-09   &   8.71$\pm$0.26   &  0.07$\pm$0.19 & 204$\pm$82   & 0.35$\pm$0.13 \\
XDOCC-10   &   8.85$\pm$0.08   &  0.40$\pm$0.17 & 655$\pm$288  & 0.29$\pm$0.12 \\
XDOCC-11   &   8.48$\pm$0.37   &  0.37$\pm$0.15 & 591$\pm$165  & 0.29$\pm$0.14 \\\hline
\end{tabular}
\end{large}
\end{table*}

Fig.~\ref{fig6} shows relatively populated to very populous Main Sequences that extend
from $\sim$ 4 up to $\sim$ 8 mag long. They do not show any clear sign of evolution,
with the exception of the presence of few stars at the Main Sequence turn-off of
XDOCC-08. Although the best representative isochrones satisfactorily reproduce these
long Main Sequences, there is no star in more evolved stellar evolutionary phases,
as we expect in open clusters with populous Main Sequences. Indeed, 
by using the Padova group web interface\footnote{http://stev.oapd.inaf.it/cgi-bin/cmd},
we generated synthetic CMDs of open clusters having the total masses, ages and metallicities 
derived for the 11 XDOCCs and found that they show 1-2 red clump stars and
very well populated Main Sequence turn-off regions. 

As \citet{bm1973} shown, a composite star field population can mimic in the CMD the
appearance of an open cluster's star sequence. Disentangling whether this is
the case of the XDOCCs' CMDs (Fig.~\ref{fig6}) is difficult to accomplish only from
the analysis of those CMDs. However, because of the minimization of likelihood functions
used to derive the astrophysical parameters, the derived bootstrapped uncertainties tell
us about the level of uniqueness of the representative solutions. Thus, under
the presence of relatively tight Main Sequences and small photometric errors, 
relatively large parameter uncertainties could suggest that different collections of
isochrones are needed to map different parts of the observed Main Sequence. 
When dealing with true open clusters, typical intrinsic dispersions of
fundamental parameters as estimated by \texttt{ASteCA} 
\citep[][and references therein]{perrenetal2022}  are $\sigma$log($t$ /yr) $\approx$ 0.12, 
and $\sigma$[Fe/H] $\approx$ 0.21 dex.
Table~\ref{tab2} shows that age uncertainties are notably large in most of the cases,
suggesting that a range of ages represent the fitted Main Sequences. The resulting metallicities
are also notably larger than the values of open clusters that follow the metallicity gradient
of the Milky Way disk. We used the Galactocentric distances of XDOCCs computed by 
\citet{jaehnigetal2021}, the recently age-metallicity-Galactocentric position relationship 
derived by \citet{magrinietal2022} from the {\it Gaia}-ESO survey \citep{gilmoreetal2012}, and ages and 
metallicities of Table~\ref{tab2}, to confirm that all XDOCCs fall outside the relationship for
open clusters. Finally, the resulting binary fractions (see Table~\ref{tab2}) generated by
\texttt{ASteCA} are relatively high. This happens when the wider broadness of the Main Sequence
of a composite star field population is assumed to be the Main Sequence of a star cluster.

\subsection{Spectroscopic data}

Chemical abundances and radial velocities for individual stars are also
helpful in order to disentangle whether they belong to the field population or to
a stellar aggregate. We took advantage of the Sloan Digital Sky Survey (SDSS) IV DR17, 
particularly the APOGEE-2 database \citep{blantonetal2017,ahumadaetal2020}, 
to retrieve this information. The spectral parameters provided in the APOGEE-2 database
were obtained using the APOGEE Stellar Parameters and Chemical Abundance Pipeline 
\citep[ASPCAP;][]{garciaperezetal2016} and were accessed employing the
following Structured Query Language (SQL) query to the SDSS database 
server\footnote{http://skyserver.sdss.org/dr17/SearchTools/sql} :\\

\texttt{SELECT TOP 100}

\texttt{s.apogee$\_$id,s.ra,s.dec,s.glon,s.glat,s.snr,}

\texttt{s.vhelio$\_$avg,s.verr,a.teff,a.teff$\_$err,a.logg,}

\texttt{a.logg$\_$err,a.m$\_$h,a.m$\_$h$\_$err,a.alpha$\_$m,}

\texttt{a.alpha$\_$m$\_$err}

\texttt{FROM apogeeStar as s}

\texttt{JOIN aspcapStar a on s.apstar$\_$id = a.apstar$\_$id}

\texttt{JOIN dbo.fGetNearbyApogeeStarEq(RA,DEC,20) as}

\texttt{ near on a.apstar\_id=near.apstar\_id}

\texttt{WHERE (a.aspcapflag \& dbo.fApogeeAspcapFlag}

\texttt{('STAR\_BAD')) = 0}\\

We then cross-matched the retrieved APOGEE-2 data sets for the 11 XDOCCs
with the corresponding {\it Gaia} data sets using RA and Dec coordinates
as matching variables and Astropy Project tools\footnote{http://www.astropy.org}.
We found only one star in common for XDOCCs~10 (2057948152709390848 ({\it Gaia} DR2 name) ==
2M20223036+3712003 (APOGEE-2 name)) with $T_{eff}$ = (16187.49$\pm$647.53)$^o$K,
log($g$)=4.252$\pm$0.082, and radial velocity = (-18.41$\pm$2.43) km/s. No metallicity
information is available. By using
the derived age, distance, and interstellar absorption of XDOCCs~10 \citep{jaehnigetal2021}, 
we found by interpolation into the corresponding theoretical isochrone that the
{\it Gaia} DR2 magnitude and colour resulted to be $G$=14.16$\pm$0.30 mag and $BP-RP$ = 
1.38$\pm$0.15 mag, respectively. These values differ significantly from the observed ones, 
namely, $G$ = 14.654 mag and $BP-RP$ = 2.048 mag. Therefore, we concluded that the star
is not at the mean distance of XDOCC~10, although its position in the CMD suggests
otherwise. This result points to the need of further spectroscopic data to assess on the
physical nature of the 11 XDOCCs. We also note that the individual {\it Gaia} DR2 
parallax uncertainties of the stars selected as members of these objects are still large 
as to secure a reliable analysis of their distances, and hence a 3D structural study of them. 
With accurate parallaxes, the 3D XDOCCs' dimensions could be estimated and from them,
a comparison with the known size range of real open clusters could be carried out.

\section{Concluding remarks}

The number of open cluster candidates identified since recent time from the 
availability of public databases and computer-based searching techniques is steadily 
increasing. Likewise, these data sets and analysis methods have been helpful
to improve the accuracy of open clusters' fundamental parameters that are
re-determined from an homogeneous basis. The outcomes of this promising
effort certainly help us to improve our knowledge of the Milky Way open
cluster system, and hence to better understand the formation and evolution
of the disc of our Galaxy. 

Recently, \citet{jaehnigetal2021} used {\it Gaia} DR2 data sets to determine the
astrophysical properties of 420 known open clusters, 
by employing extreme deconvolution gaussian mixture models.
They also pointed out that other previously unknown 11 open cluster candidates 
could be populating the searched sky regions. The identification of these
new candidates mainly relies on stellar overdensities detected in the
Vector Point diagram, whose stars are distributed in CMDs following the appearance of 
those typical of open clusters; the distribution of parallaxes also turned out to be at
first glance compatible with  stars being at a similar mean distance. Nevertheless,  
 \citet{jaehnigetal2021} stressed that further investigations are necessary in order 
to confirm their physical nature.

By using the same data sets, we carried out a thorough analysis of these
new open cluster candidates (XDOCCs), from which we identified a number of 
astrophysical aspects that do not fully agree with our present knowledge
of the open cluster population. These considerations come from our
independent estimates of age, reddening, metallicity, total mass, and binary 
fraction of the 11 XDOCCs, considering them as real physical stellar systems.
These fundamental parameters were derived by disentangling their
intrinsic dispersions from the photometric data sets uncertainties. We obtained
astrophysical properties whose associated dispersions are much larger than 
those usually obtained for open clusters using the same procedure.
Such large dispersions are typical of fundamental parameters of
composite star field populations.

The apparent distribution of stars in the sky, their projected
stellar number density profiles, and the relationship between their
masses and projected radii support also that, considered all together,
the conclusion that the XDOCCs unlikely are real physical systems.
We think that this finding highlights the need of gathering more than one
criterion when searching for open cluster candidates. Here we show that
a stellar overdensity  in the proper motion space is not enough to
conclude on the existence of a stellar aggregate. By relying on a
multidimensional approach (3D positions, 3D motions, metallicity,
CMD features, etc) more confidence open cluster candidates will be uncovered.
Nevertheless, because of the clearly observed gathering of stars in the
Vector Point diagram and some hint for similar distances within the
{\it Gaia} DR2 parallax uncertainties, the XDOCCs  could be
considered like sparse groups of stars.


\section*{Acknowledgements}
We thank the referee for the thorough reading of the manuscript and
timely suggestions to improve it.

This work has made use of data from the European Space Agency (ESA) mission
{\it Gaia} (\url{https://www.cosmos.esa.int/gaia}), processed by the {\it Gaia}
Data Processing and Analysis Consortium (DPAC,
\url{https://www.cosmos.esa.int/web/gaia/dpac/consortium}). Funding for the DPAC
has been provided by national institutions, in particular the institutions
participating in the {\it Gaia} Multilateral Agreement.

This work made use of Astropy: a community-developed 
core Python package and an ecosystem of tools and resources for astronomy 
\citep{astropy2013, astropy2018, astropy2022}.

\section{Data availability}

Data used in this work are available upon request to the author.










\bsp	
\label{lastpage}
\end{document}